\documentclass[runningheads]{llncs}
\usepackage[T1]{fontenc}
\usepackage{amsmath,amssymb,amsfonts}
\usepackage{float}
\usepackage{algorithmic}
\usepackage{graphicx}
\usepackage{textcomp}
\usepackage{xcolor}
\usepackage[numbers]{natbib} 
\usepackage{multirow}
\begin{document}
\title{BlastDiffusion: A Latent Diffusion Model for
Generating Synthetic Embryo Images to Address
Data Scarcity in In Vitro Fertilization}
\titlerunning{BlastDiffusion: Latent Diffusion for Synthetic Embryo Images in IVF}
%
\author{Alejandro Golfe\inst{1,}$^*$\orcidID{0000-0001-5095-9904} \and
Natalia P. García-de-la-Puente \inst{1}\orcidID{0009-0009-9704-9102} \and
Adrián Colomer\inst{1}\orcidID{0000-0002-7616-6029} \and Valery Naranjo\inst{1}\orcidID{0000-0002-0181-3412}}

\authorrunning{F. Author et al.}
%
\institute{\textsuperscript{1} Instituto de Investigación sobre Tecnología Centrada en el Ser Humano (Human-Tech), Universidad Politécnica de Valencia, España \\ 
$^*$ Correspondng author: Alejandro Golfe (algolsan@upv.es)} 

\maketitle              
\begin{abstract}
Accurately identifying oocytes that progress to the blastocyst stage is crucial in reproductive medicine, but the limited availability of annotated high-quality embryo images presents challenges for developing automated diagnostic tools. To address this, we propose BlastDiffusion, a generative model based on Latent Diffusion Models (LDMs) that synthesizes realistic oocyte images conditioned on developmental outcomes. Our approach utilizes a pretrained Variational Autoencoder (VAE) for latent space representation, combined with a diffusion process to generate images that distinguish between oocytes that reach the blastocyst stage and those that do not. When compared to Blastocyst-GAN, a GAN-based model we trained for this task, BlastDiffusion achieves superior performance, with a global Frechet Inception Distance (FID) of 94.32, significantly better than Blastocyst-GAN’s FID of 232.73. Additionally, our model shows improvements in perceptual (LPIPS) and structural (SSIM) similarity to real oocyte images. Qualitative analysis further demonstrates that BlastDiffusion captures key morphological differences linked to developmental outcomes. These results highlight the potential of diffusion models in reproductive medicine, offering an effective tool for data augmentation and automated embryo assessment.


\end{abstract}

\section{Introduction}

Reliable identification of oocytes that progress to the blastocyst stage is vital in reproductive biology and clinical medicine. Blastocysts represent a key phase in embryonic development, as this is when embryos are most likely to implant successfully in the recipient's uterus, increasing the chances of a viable pregnancy \cite{gardner1997culture}. Furthermore, analyzing embryos at this stage provides valuable insights into developmental quality and helps select the best embryos for transfer in vitro fertilization (IVF) treatments \cite{gardner1998prospective}.

However, there is a significant scarcity of annotated datasets containing high-quality images of embryos at the blastocyst stage, which limits both research progress and clinical applications of computer vision algorithms. The absence of consistent and standardized labelling of the ground truth in training data hinders artificial intelligence (AI) models' reliability and predictive capability in embryo selection \cite{salih2023embryo}. In this context, generating synthetic images could provide a solution to expand the available data, allowing the training of more robust models for oocyte identification and classification at this crucial stage.

Generative models have evolved significantly in recent years, from Variational Autoencoders (VAEs) \cite{kingma2013auto}, which capture latent data distributions, to Generative Adversarial Networks (GANs) \cite{goodfellow2014generative}, which enable the creation of realistic images. More recently, diffusion models, which rely on transforming data through a series of noise and restoration steps, have demonstrated exceptional performance in generating high-quality images \cite{croitoru2023diffusion}. Due to their ability to generate complex and realistic images, these models present a promising tool for creating images of embryos at the blastocyst stage, contributing to the expansion of annotated datasets and improving the accuracy of automated diagnostic systems. The contributions of this work can be summarized as follows:  
 
\begin{itemize}
    \item We introduce the first Latent Diffusion Model (LDM) that generates synthetic oocyte images conditioned on their capacity to reach the blastocyst stage, directly addressing the scarcity of annotated data for research and clinical applications.
    \item By leveraging diffusion models to produce high‑quality synthetic images, this work expands the data available for embryo‑selection research and clinical practice in IVF, illustrating the promise of advanced generative models in biomedical applications.
    \item We perform benchmarking against Blastocyst-GAN, a GAN-based architecture we trained for oocyte image generation, highlighting the superior performance of our diffusion-based approach in key metrics like FID, LPIPS, and SSIM.
\end{itemize}

\section{Related work}

Accurately identifying and selecting blastocysts in reproductive medicine have been extensively studied, with various approaches proposed to improve embryo assessment. Traditional embryo evaluation methods rely on morphological grading systems \cite{gardner1997culture}, which, while effective, are subjective and prone to interobserver variability \cite{martinez2017inter}. To address these limitations, automated systems using machine learning and deep learning have been explored to enhance embryo selection in IVF treatments \cite{hew2024artificial}.

The availability of well-labelled embryo images is critical for training robust classification and predictive models, yet existing datasets are limited in size and diversity \cite{hew2024artificial}. To overcome this challenge, data augmentation techniques, including synthetic image generation, have been proposed to expand training datasets and improve model performance \cite{wang2024enhance}. However, traditional data augmentation techniques, such as colour jittering, flips, rotations, and random cropping, are limited in capturing complex morphological variations. These methods alter pixel-level features without introducing new structural patterns, which may be crucial for modelling fine-grained biological differences.

Generative models have played a key role in synthetic data generation. LDMs have emerged as a powerful alternative for image synthesis, offering a more computationally efficient approach that operates in a lower-dimensional latent space \cite{herron2024latent}. These models iteratively transform noise into detailed images through a learned denoising process, enabling the generation of highly realistic and diverse datasets \cite{croitoru2023diffusion}. LDMs have been successfully applied in various biomedical imaging tasks, including histopathology \cite{yellapragada2024pathldm} and embryo generation \cite{presacan2024embryo}.

Recent developments in diffusion models have emphasized the integration of cross-attention layers, which allow the generative process to condition on additional input information such as class labels, textual descriptions, or clinical metadata. Cross-attention improves the model’s ability to focus on relevant spatial and semantic features, making it particularly effective for conditional image generation tasks. In biomedical imaging, this facilitates the synthesis of images that reflect subtle biological variations linked to specific phenotypes or clinical outcomes \cite{rombach2022high, dhariwal2021diffusion}.

In the context of reproductive medicine, applying diffusion models with cross-attention layers enables conditioning synthetic oocyte images on developmental outcomes—specifically, whether an oocyte develops into a blastocyst or not. This conditional generation is crucial because oocytes that do or do not reach the blastocyst stage exhibit distinct morphological features that should be accurately represented in the synthetic data to better reflect biological variability.

Given their superior image generation capabilities, diffusion models represent a promising approach to address dataset limitations in embryo classification. Previous studies have demonstrated that augmenting training data with synthetic images generated by diffusion or other generative models can significantly enhance classification performance and model robustness in biomedical imaging tasks \cite{golfe2023progleason, oh2023diffmix}. Unlike previous work focusing on direct embryo generation \cite{presacan2024embryo}, our approach leverages latent diffusion models (LDMs) to generate synthetic oocyte images conditioned on developmental outcome information, specifically whether they reach the blastocyst stage or not. To the best of the authors' knowledge, this is the first work to apply LDMs for conditional oocyte image generation based on developmental viability.

\section{Materials}

\subsection{Database}
The protocol and procedures for oocyte analysis were approved by the Institutional Review Board (IRB reference 2303-VLC-035-MM), which oversees database analyses and clinical IVF research procedures at IVI (Instituto Valenciano de Infertilidad) and RMA (Reproductive Medicine Associates), now known as IVIRMA Global. This retrospective study examined 2,217 oocyte images captured before intracytoplasmic sperm injection (ICSI) using a Basler acA3088-57uc camera connected to a microinjection microscope. Patient data, stimulation protocols, clinical outcomes, and work environment variables were collected from the center's internal registry.

All data used in this study was pseudonymized, as the investigator receives the data without any identification. The individual responsible for data extraction retained re-identification information, ensuring a clear functional and technical separation between the researcher and data collector. The compiled data were divided into training, validation, and testing sets while ensuring that samples from the same patient did not appear in both training and testing sets simultaneously. The training and validation set comprised 90\% of the total data, with 10\% of this portion further allocated for validation. The remaining 10\% of the data was reserved for testing. In the dataset, 44.8\% of the oocytes correspond to those that reach the blastocyst stage, while 55.2\% correspond to those that do not. The partitioning was performed at the patient level to prevent data leakage and ensure a robust model evaluation.

A key preprocessing step involves isolating the region in each image containing the oocyte. To achieve this, we use a pretrained YOLO model \cite{varghese2024yolov8}, which is well-known for its effectiveness in object detection and real-time detection capabilities and facilitates its integration into oocyte analysis software. This architecture has been widely adopted in various medical imaging applications, including tumor localization, fracture detection, and histopathological analysis, demonstrating its versatility and robustness across domains \cite{soni2024yolo}. This process reduces the image dimensions from approximately 3000 × 2000 to 1500 × 1500 pixels. For retraining the YOLOv8n model, bounding boxes were manually labelled on 61 images, with a train/validation/test split of 26/24/11 \cite{garcia2024unsupervised}.

\section{Methodology}
\begin{figure*}[h]
    \centering
    \includegraphics[width=\textwidth]{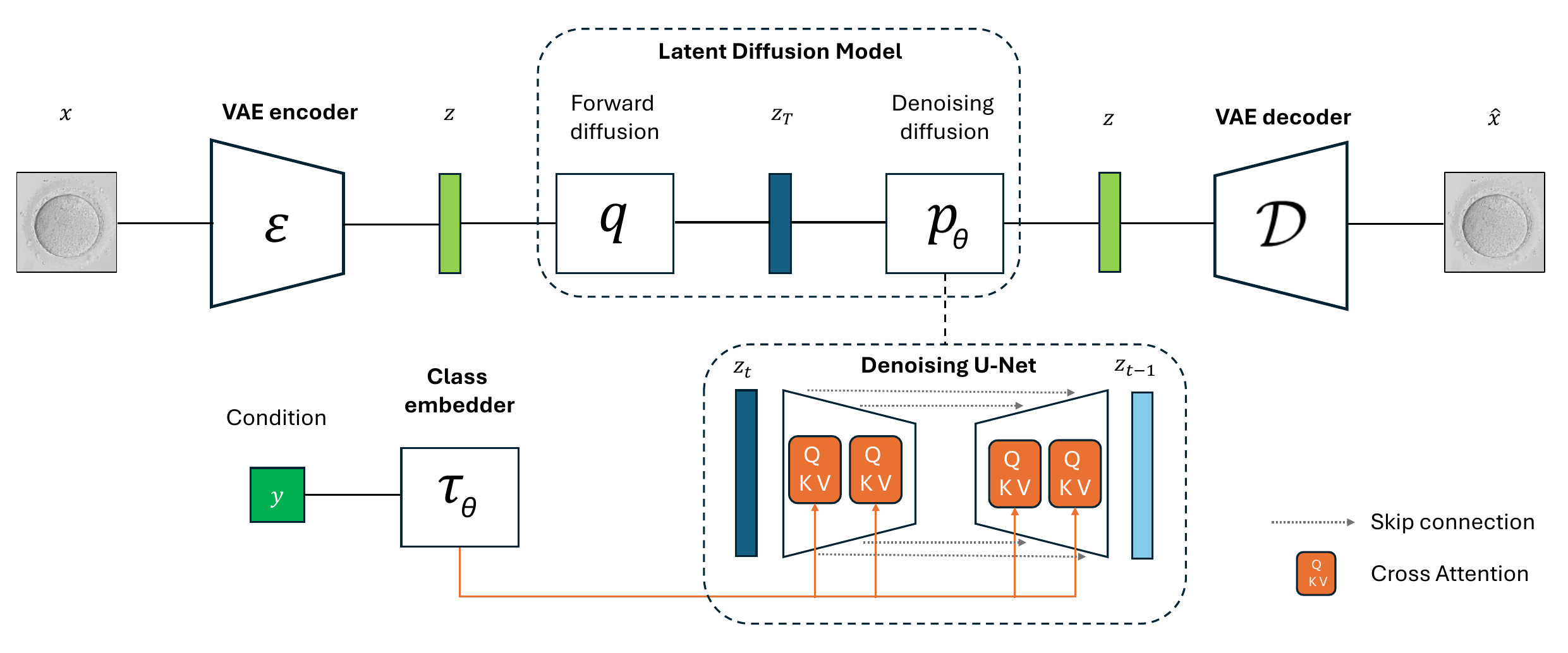}
    \caption{Overview of BlastDiffusion framework. A Variational Autoencoder (VAE) encodes the input image into a latent space, where a LDM refines noisy representations into realistic embeddings. The process is conditioned on developmental outcome, allowing the generation of biologically relevant synthetic images.}
    \label{fig:methodology}
\end{figure*}

Our approach leverages a LDM for conditional oocyte image generation. First, the VAE encoder compresses the input image into a lower-dimensional latent space, facilitating efficient processing. Then, the LDM is applied within this latent space, iteratively refining noisy representations into realistic embeddings. A class embedding mechanism conditions the generation process on whether the oocyte reaches the blastocyst stage, guiding the diffusion model toward biologically relevant image synthesis. Finally, the VAE decoder reconstructs the high-resolution image from the generated latent representation (see Fig \ref{fig:methodology}).

\subsection{BlastDiffusion Framework}

\subsubsection{Variational Autoencoder (VAE) Encoder}  
To obtain a compact and structured latent representation, we employ a pretrained Variational Autoencoder (VAE) from the Stable Diffusion framework. This VAE, originally trained on the LAION-5B dataset \cite{schuhmann2022laion}, filtered to exclude sensitive content, serves as a feature extractor. Given an input image \( x \in \mathbb{R}^{H \times W \times C} \), the VAE encoder \( \mathcal{E} \) compresses it into a lower-dimensional latent representation \( z \in \mathbb{R}^{h \times w \times c} \), where typically \( h < H \), \( w < W \), and \( c < C \). This transformation preserves the most relevant features of the data while significantly reducing computational overhead compared to operating directly in pixel space. The latent representation \( z \) is then used as input for the diffusion process.

Mathematically, the encoder \( \mathcal{E} \) can be expressed as:

\[
z = \mathcal{E}(x) \quad \text{where} \quad z \sim q_{\phi}(z|x)
\]

where \( \mathcal{E}(x) \) represents the encoder function mapping the input image \( x \) to the latent variable \( z \), and \( q_{\phi}(z|x) \) is the approximate posterior distribution of \( z \) given \( x \), parameterized by \( \phi \).

The VAE decoder \( \mathcal{D} \) reconstructs the input image from the latent representation. Given the latent variable \( z \), the VAE decoder \( \mathcal{D} \) generates the reconstructed image \( \hat{x} \in \mathbb{R}^{H \times W \times C} \), which can be expressed as:

\[
\hat{x} = \mathcal{D}(z) \quad \text{where} \quad p_{\theta}(x|z) = \mathcal{N}(\hat{x}, \sigma^2)
\]

where \( \mathcal{D}(z) \) is the decoder function that maps the latent representation \( z \) back to the image space to produce the reconstruction \( \hat{x} \), and \( p_{\theta}(x|z) \) is the likelihood of observing the image \( x \) given the latent variable \( z \), modeled as a Gaussian distribution with mean \( \hat{x} \) and variance \( \sigma^2 \).

\subsubsection{Diffusion Process}  
The diffusion model learns to generate samples by gradually corrupting a data distribution with noise and then training a neural network to reverse this process. Given a latent representation \( z_0 \) obtained from the VAE encoder, a \textit{forward diffusion process} progressively adds Gaussian noise through a predefined schedule, producing a sequence of latent variables \( z_1, z_2, \dots, z_T \). This process follows the formulation:  

\begin{equation}
    q(z_t \mid z_{t-1}) = \mathcal{N}(z_t; \sqrt{\alpha_t} z_{t-1}, (1 - \alpha_t) I)
\end{equation}

where \( q(z_t \mid z_{t-1}) \) represents the transition probability at time step \( t \). The term \( \alpha_t \) is a noise scheduling coefficient that controls the amount of noise added at each step, ensuring a smooth transition from structured data to pure noise. Here, \( I \) is the identity matrix of appropriate size, which is used to model the covariance structure of the added noise, ensuring that the noise is independent across the dimensions of \( z_t \).

During training, the model learns to approximate the reverse process, denoted as:  

\begin{equation}
    p_\theta(z_{t-1} \mid z_t) = \mathcal{N}(z_{t-1}; \mu_\theta(z_t, t), \Sigma_\theta(z_t, t))
\end{equation}

where \( p_\theta(z_{t-1} \mid z_t) \) models the probability distribution of recovering \( z_{t-1} \) given \( z_t \). The function \( \mu_\theta(z_t, t) \) represents the predicted mean of the denoised latent variable, while \( \Sigma_\theta(z_t, t) \) is the learned variance. The parameters \( \theta \) are optimized to minimize the divergence between the true reverse process and its approximation.  

\subsubsection{Conditioning Mechanism}  
To guide image synthesis toward meaningful biological relevance, we incorporate a conditioning mechanism that introduces information about whether an oocyte develops into a blastocyst. The diffusion model is conditioned on an embedding vector \( y \) that encodes this classification (blastocyst or non-blastocyst). This embedding is processed through a learnable transformation \( \tau_\theta \), yielding a conditioning vector:  

\[
c = \tau_\theta(y)
\]

which is then injected into the diffusion process via \textit{cross-attention layers}, as illustrated in Fig.~\ref{fig:methodology}. The conditional reverse process is formulated as:  

\begin{equation}
    p_\theta(z_{t-1} \mid z_t, c) = \mathcal{N}(z_{t-1}; \mu_\theta(z_t, c, t), \Sigma_\theta(z_t, c, t))
\end{equation}

where \( \mu_\theta(z_t, c, t) \) and \( \Sigma_\theta(z_t, c, t) \) now depend on both the noisy latent variable \( z_t \) and the conditioning vector \( c \). This conditioning enables the model to generate synthetic oocyte images that are representative of the specified developmental outcome, thereby augmenting datasets in a targeted manner.

\subsection{Blastocyst-GAN Framework}
The Blastocyst-GAN is a conditional generator based on the ProGleason-GAN architecture \cite{golfe2023progleason}. Although the original ProGleason-GAN was trained on prostate histology patches, the Blastocyst-GAN model has been trained from scratch using our oocyte dataset. This adaptation enables the generation of synthetic samples conditioned on whether the oocytes develop into viable blastocysts or not. This tailored training ensures that the model captures oocytes' unique characteristics and developmental outcomes.

We used this newly trained Blastocyst-GAN model as a comparison baseline to evaluate our proposed method more robustly. By comparing the results of our method with the outcomes generated by this model, we can assess the strengths and weaknesses of our approach, offering a comprehensive evaluation.

In the original ProGleason-GAN model, the generator was conditioned on the Gleason grade in prostate tissue patches. However, for our study, this mechanism was modified so that the synthesis process is conditioned on the developmental status of oocytes (viable or non-viable blastocyst).

The ProGAN \cite{karras2017progressive} architecture enables the model to learn high-level image features and refine them as training progresses. The generator and the discriminator are provided with information about the specific condition of the oocyte (whether it results in a viable or non-viable blastocyst), allowing for conditional image synthesis without needing to incorporate a specific term in the loss function.

The training process is conducted in multiple stages, beginning with low-resolution patches of 4x4 pixels and progressively increasing to a final resolution of 512x512 pixels. Additionally, the model employs a technique called fade-in, which smoothly transitions between different resolutions, helping to stabilize the training process. Techniques such as minibatch standard deviation and pixel normalization are also used to improve the quality of the generated images.

The model is trained using the Wasserstein GAN with Gradient Penalty (WGAN-GP) loss function \cite{gulrajani2017improved}, which minimizes the discrepancy between real and synthetic data distributions, thereby improving the quality of the generated synthetic oocyte images.

\section{Experiments and Results}
\subsection{Experimental setup}

\subsubsection{Training Configuration}

The BlastDiffusion model was trained using the Adam optimizer with a learning rate of 0.00005. A batch size of 24 was used, and the training was carried out for 4,000 epochs. The latent space parameters for the denoising U-net included downsampling channels of [64, 128, 256, 512] and the middle layers of [512, 256]. The time embedding dimension was set to 512. The upsampling process was carried out in two stages, using layers that mirror the downsampling process for efficient reconstruction. A conditional dropout probability of 0.2 was applied to introduce diversity in the generated images, and the training data was augmented using random horizontal flips (p=0.5) and random vertical flips (p=0.3). Finally, a perceptual loss weight of 1 was used to ensure high-quality image generation.

\subsubsection{Implementation}

The experiments were conducted on an NVIDIA DGX A100 system with six NVIDIA A100 GPUs, each having 40 GB of HBM2 memory. Multi-GPU training was utilized to accelerate model training. The environment used was PyTorch 2.1.2 and Python 3.10.
    
\subsubsection{Evaluation Metrics}

We used multiple metrics to evaluate the synthetic images to assess their quality and similarity to real images. The primary metric employed was the Frechet Inception Distance (FID) \cite{heusel2017gans}, which measures the distance between two distributions of images: real and generated images. It uses features extracted from a pre-trained Inception v3 model \cite{szegedy2016rethinking}, providing a high-dimensional image embedding.  

Formally, FID is defined as the Frechet distance between the Gaussian distributions fitted to these embeddings:

\[
FID = \left\| \mu_r - \mu_g \right\|_2^2 + \text{Tr}(\Sigma_r + \Sigma_g - 2(\Sigma_r \Sigma_g)^{1/2})
\]

Where \(\mu_r\) and \(\Sigma_r\) represent the mean and covariance of the real images' embeddings, while \(\mu_g\) and \(\Sigma_g\) represent the mean and covariance of the generated images' embeddings. The term \(\left\| \cdot \right\|_2\) denotes the L2 norm, which measures the Euclidean distance between the means of the two distributions. The trace operator denoted as \(\text{Tr}\), is applied to the sum of the covariance matrices of both real and generated images. Lower FID values indicate that the generated images are closer to the real images in terms of both distribution and visual quality.

In addition to FID, we also computed the Learned Perceptual Image Patch Similarity (LPIPS) metric \cite{johnson2016perceptual}, which evaluates the perceptual similarity between real and synthetic images based on deep neural network activations. Lower LPIPS values indicate higher perceptual similarity between the images.

Furthermore, the Structural Similarity Index (SSIM) was used to evaluate the structural similarity between real and generated images. This metric compares the luminance, contrast, and structural information between images.

\subsection{Quantitative evaluation}

The results presented in Table~\ref{tab:results} show that our proposed method, BlastDiffusion, outperforms Blastocyst-GAN across all evaluated metrics. Specifically, BlastDiffusion achieves a significant reduction in FID across all categories, indicating a higher similarity to real data. For the LPIPS metric, it also obtains lower values, with a particularly notable improvement in the blastocyst class, suggesting that the generated images are perceptually closer to real ones. Additionally, the SSIM values demonstrate better structural preservation in BlastDiffusion-generated images, with consistent improvements across all categories, especially in the blastocyst class. These results confirm that BlastDiffusion generates images of higher quality, more realistic and structurally coherent than Blastocyst-GAN, establishing itself as a superior alternative for oocyte generation.  

\begin{table}[]
\centering
\begin{tabular}{c@{\hspace{8pt}}c@{\hspace{10pt}}c@{\hspace{8pt}}c@{\hspace{8pt}}c}
\textbf{Method}                  & \textbf{\centering Class}    & \multicolumn{1}{c}{\textbf{FID $\downarrow$}} & \textbf{LPIPS $\downarrow$} & \textbf{SSIM $\uparrow$} \\
\hline
\multirow{3}{*}{Blastocyst-GAN}   & \centering Non-Blastocyst    & 223.588  & 0.392 $\pm$ 0.054  & 0.312 $\pm$ 0.056  \\
                                  & \centering Blastocyst       & 278.581             & 0.397 $\pm$ 0.064  & 0.308 $\pm$ 0.052  \\
                                  & \centering Total           & 232.733                & 0.394 $\pm$ 0.059  & 0.310 $\pm$ 0.054  \\
\hline  
\multirow{3}{*}{BlastDiffusion}    & \centering Non-Blastocyst    & \textnormal{\textbf{101.166}}  & \textnormal{\textbf{0.2745 $\pm$ 0.0540}}  & \textnormal{\textbf{0.3913 $\pm$ 0.0506}} \\
                                  & \centering Blastocyst       & \textnormal{\textbf{105.503}}              & \textnormal{\textbf{0.2659 $\pm$ 0.0511}}  & \textnormal{\textbf{0.3863 $\pm$ 0.0572}} \\
                                  & \centering Total            & \textnormal{\textbf{94.318}}              & \textnormal{\textbf{0.2877 $\pm$ 0.0461}}  & \textnormal{\textbf{0.4462 $\pm$ 0.0534}} \\
\end{tabular}
\vspace{5pt}
\caption{Comparison of different methods for FID, LPIPS, and SSIM metrics across different classes.}
\label{tab:results}
\end{table}

The limited dataset size is a key factor in interpreting these results, which challenges capturing fine details and full variability within each class. Despite this, the results highlight the model's ability to conditionally generate images, effectively distinguishing embryos that reach the blastocyst stage from those that do not.

\subsection{Qualitative Evaluation}

\begin{figure}[h]
    \centering
    \includegraphics[width=1\textwidth]{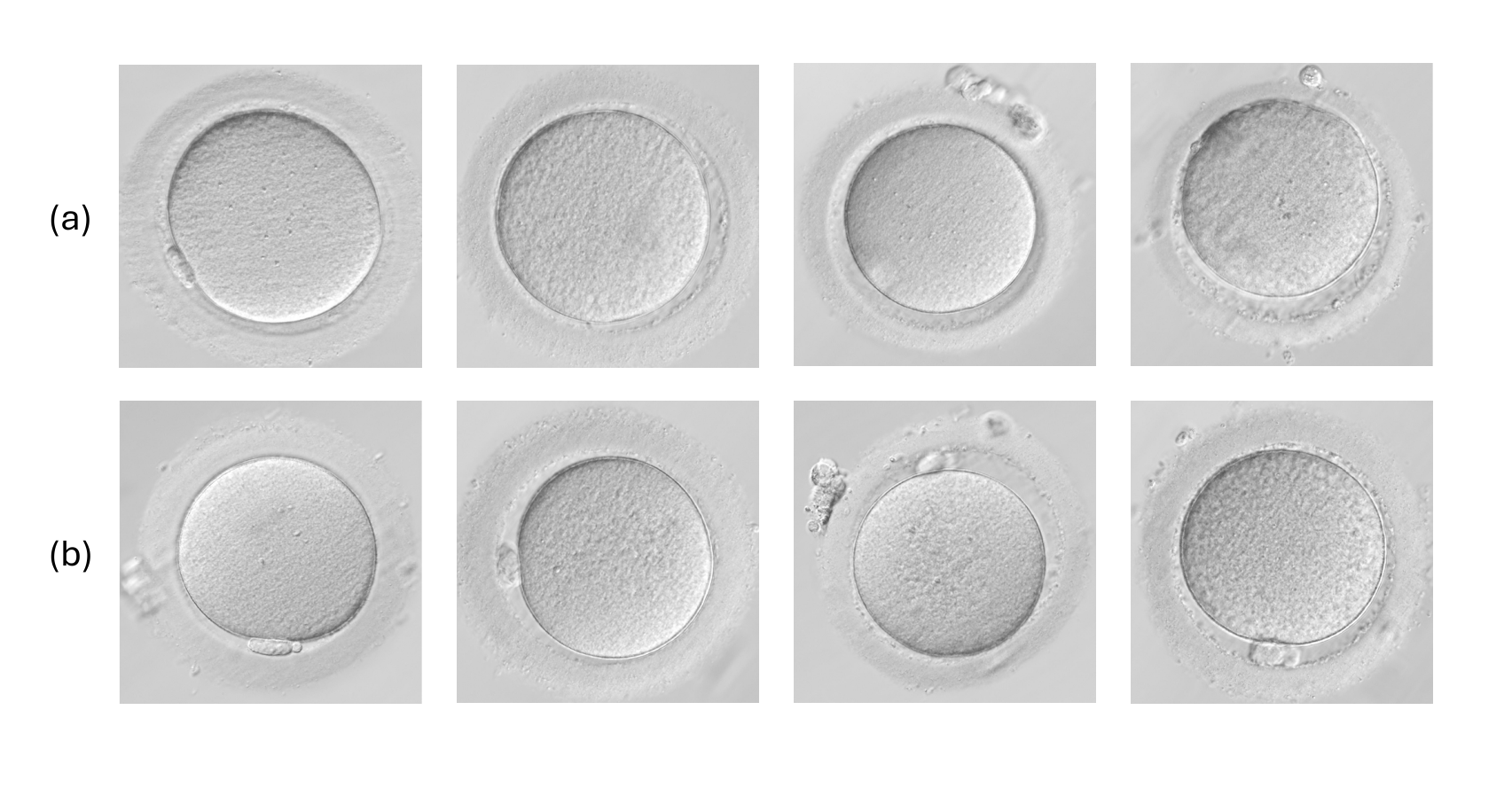}
    \caption{Real oocyte images. The first row (a) corresponds to oocytes that did not reach the blastocyst stage, while the second row (b) shows oocytes that successfully developed into blastocysts.}
    \label{fig:real_images}
\end{figure}

\begin{figure}[H]
    \centering
    \includegraphics[width=1\textwidth]{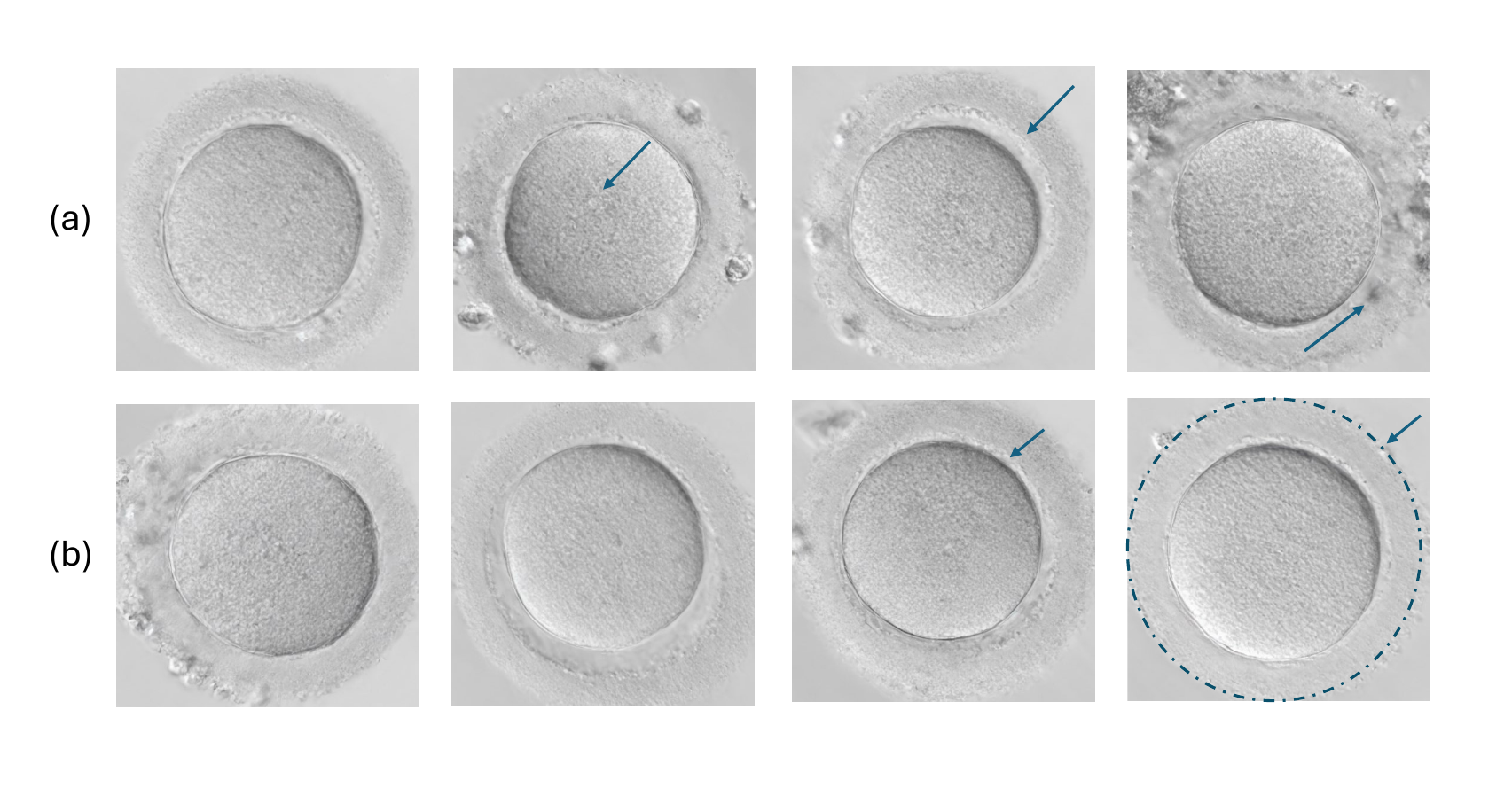}
    \caption{Generated oocytes using BlastDiffusion. The first row (a) consists of generated oocytes that did not reach the blastocyst stage, while the second row (b) represents those predicted to develop into blastocysts. Specific features are highlighted with arrows.}
    \label{fig:diffusion_images}
\end{figure}

\begin{figure}[h]
    \centering
    \includegraphics[width=1\textwidth]{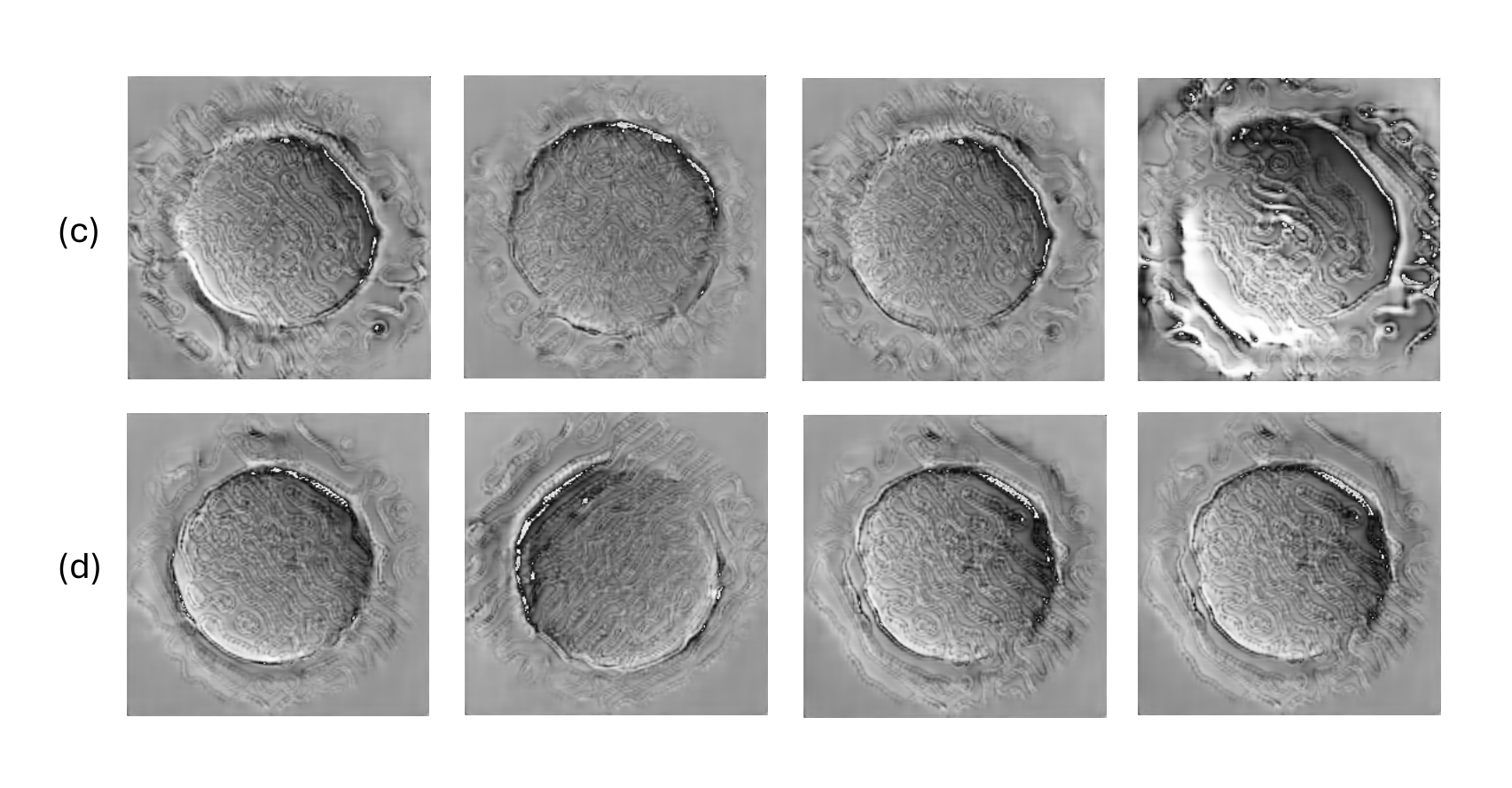}
    \caption{Generated oocytes using Blastocyst-GAN. The first row (a) represents generated oocytes that did not reach the blastocyst stage, while the second row (b) corresponds to those predicted to develop into blastocysts.}
    \label{fig:gan_images}
\end{figure}

To assess the quality of the images generated by Blastocyst-GAN and BlastDiffusion, we compare them visually against real oocytes. Figure \ref{fig:real_images} shows real oocytes, Figure \ref{fig:gan_images} shows oocytes generated by Blastocyst-GAN and Figure \ref{fig:diffusion_images} oocytes generated by BlastDiffusion. In each set, the first row corresponds to oocytes that do not reach the blastocyst stage, while the second row represents those that do.

From a biological perspective, the model must differentiate between oocytes that reach the blastocyst state and oocytes that do not. In the images generated by BlastDiffusion (Fig. \ref{fig:diffusion_images}), a clearer separation between these two categories is observed. Row (a) oocytes display characteristics often associated with lower developmental potential, including an irregular and granular cytoplasm (as highlighted in the second image of row a), and a zona pellucida that appears less smooth and may exhibit adherent debris within the perivitelline space (noticeable in the fourth image of row a). Furthermore, the perivitelline space in row (a) sometimes appears wider and less uniform (observe the third image of row a).

In contrast, row (b) oocytes, which reached the blastocyst stage, show a smooth and homogeneous cytoplasm (evident in the third image of row b), a clean perivitelline space often devoid of visible debris (clearly seen in all images of row b, especially the fourth), and a seemingly more uniformly structured zona pellucida (as indicated in the fourth image of row b). In particular, the oocytes in the second row exhibit a more homogeneous internal structure and a more distinct and regular zona pellucida, characteristics also present in the real images (Fig. \ref{fig:real_images}). In contrast, Blastocyst-GAN (Fig. \ref{fig:gan_images}) tends to generate less differentiated patterns, suggesting a lower capacity to capture key biological differences.

BlastDiffusion demonstrates a notable improvement in visual fidelity compared to Blastocyst-GAN. While the images generated by Blastocyst-GAN exhibit noticeable artefacts and unnatural textures, BlastDiffusion produces oocytes with more defined structures and a smoother texture that closely resembles real samples. In particular, peripheral cellular structures and the zona pellucida appear to be more naturally defined in the images generated by BlastDiffusion.

In conclusion, the conditional synthesis of BlastDiffusion shows significant improvements in oocyte generation, both in terms of visual realism and the accurate representation of relevant biological features. These results indicate that BlastDiffusion produces more realistic images and is a more reliable tool for computational embryology studies. Although the limited size of the data set may restrict the model's ability to learn fine-grained morphological features associated with blastocyst formation, BlastDiffusion still demonstrates a remarkable capacity to capture intrinsic class characteristics despite data limitations, highlighting its potential for applications in oocyte analysis. One possible explanation for its superior performance compared to Blastocyst-GAN is the architectural difference between the models: while BlastDiffusion leverages diffusion-based generation with cross-attention mechanisms that enable finer control over conditional synthesis, Blastocyst-GAN is based on a progressive growing GAN framework, which may be less effective in capturing complex, high-resolution biological details under limited data conditions. These observations highlight the potential of conditional generative models in reproductive biology and the importance of further improving dataset diversity.

\section{Conclusion}

BlastDiffusion, a conditional latent‑diffusion model trained with a pretrained VAE backbone, synthesizes oocyte images conditioned on later blastocyst formation. Despite using only binary class labels, it achieves markedly superior realism and biological relevance over our GAN baseline, improving FID from 232.7 to 94.3 and surpassing it in LPIPS and SSIM while faithfully reproducing morphological cues that distinguish oocytes that reach the blastocyst stage from those that do not.

While our model relies on a compact dataset and uses only binary class‑label conditioning, these choices mark clear opportunities for growth rather than shortcomings. Scaling the dataset and complementing the class labels with richer signals—such as text annotations or molecular markers—should further sharpen image specificity and widen the spectrum of oocyte phenotypes the model can represent.

Future work should enlarge and diversify the training corpus and enrich conditioning with multimodal biological metadata. Crucially, synthetic images ought to be validated by an independent embryo‑quality classifier and, eventually, through prospective clinical evaluation to confirm their utility in IVF practice.

\section*{Acknowledgment}
This work has received funding from the Spanish Ministry of Economy and Competitiveness through the project PID2022-140189OB-C21 (ASSIST). The work of Natalia P. García de la Puente was supported by the grant PID2022-140189OB-C21 funded by MICIU/AEI/10.13039/ 501100011033 ERDF/UE and FSE+.

\bibliographystyle{IEEEtran} 
\bibliography{cas-refs}
\end{document}